\documentclass[twocolumn,amsmath,amssymb,footinbib,superscriptaddress]{revtex4-1}

\usepackage{graphicx}
\usepackage{dcolumn}
\usepackage{bm}
\usepackage{color}
\usepackage{amsmath}
\usepackage[colorlinks=true,citecolor=blue,urlcolor=blue,linkcolor=blue]{hyperref}

\begin{document}

\title{Sub-picosecond acoustic pulses at buried GaP/Si interfaces}

\author{Kunie Ishioka}
\email{ishioka.kunie@nims.go.jp}
\affiliation{National Institute for Materials Science, Tsukuba, 305-0047 Japan}

\author{Avinash Rustagi}
\affiliation{Department of Physics, University of Florida, Gainesville, FL 32611 USA}

\author{Andreas Beyer}
\affiliation{Philipps-University, Material Sciences Center and Faculty of Physics, D-35032 Marburg, Germany}

\author{Wolfgang Stolz}
\affiliation{Philipps-University, Material Sciences Center and Faculty of Physics, D-35032 Marburg, Germany}

\author{Kerstin Volz}
\affiliation{Philipps-University, Material Sciences Center and Faculty of Physics, D-35032 Marburg, Germany}

\author{Ulrich H{\"o}fer}
\affiliation{Philipps-University, Material Sciences Center and Faculty of Physics, D-35032 Marburg, Germany}

\author{Hrvoje Petek}
\affiliation{Department of Physics and Astronomy, University of Pittsburgh, Pittsburgh, PA 15260, USA}

\author{Christopher J. Stanton}
\affiliation{Department of Physics, University of Florida, Gainesville, FL 32611 USA}

\date{\today}

\begin{abstract}

We report on the optical generation and detection of ultrashort acoustic pulses that propagate in three-dimensional semiconductor crystals.  Photoexcitaiton of lattice-matched GaP layers grown on Si(001) gives rise to a sharp spike in transient reflectivity due to the acoustic pulse generated at the GaP/Si interface and detected at the GaP surface and vice versa.  The extremely short width of the reflectivity spike, 0.5 ps, would translate to a spatial extent of 3 nm or 10 atomic bilayers, which is comparable with the width of the intermixing layer at the GaP/Si interface.  The reflectivity signals are also modified by quasi-periodic Brillouin oscillations of GaP and Si arising from the acoustic pulses during the propagation in the crystals. The present results demonstrate the potential application of the simple optical pump-probe scheme in the nondestructive evaluation of the buried semiconductor interface quality.

\end{abstract}

\pacs{78.47.jg, 63.20.kd, 78.30.Fs}
\maketitle


Photoexcitation of metals and semiconductors with ultrashort laser pulses can generate strain or shear pulses which propagate into  the bulk crystals or along the surfaces  \cite{Thomsen86}.  They are induced predominantly through sudden lattice heating (thermoelastic effect) in metals and via deformation potential coupling with (and piezoelectric screening by) photoexcited carriers in semiconductors.    These acoustic pulses, widely known as coherent acoustic phonons,  have been studied extensively for potential applications in determination of the mechanical properties of solids \cite{Ogi07, Lomonosov12}, ultrafast optical control of piezoelectric effect \cite{Sun01} and magnetism \cite{Kim15}, development of nanoplasmonic resonators \cite{OBrien14} and terahertz polaritonics \cite{Koehl01}.  

Coherent acoustic phonons can also be a useful tool in nano-seismology and nano-tomography for characterizing buried interfaces and defective layers, objects hidden under the surfaces, molecules adsorbed on surface and local strains \cite{Wright92, Tas98, Hettich12, Kashiwada06, Dehoux09}.  One way to realize ultrashort acoustic pulses for higher spatial resolution in these applications is to generate surface acoustic waves (SAWs) using one- or two-dimensional metallic nano-gratings.  The shortest wavelength of SAWs achieved so far was 45 nm, determined by the periodicity of the grating \cite{Li12}.  Another approach is to confine coherent acoustic phonon pulses into sub-micron scale objects.  The shortest acoustic pulse actually measured, with a wavelength of 150 nm, was achieved by focusing an acoustic pulse inside a metal-coated silica fiber \cite{Dehoux16}.  However, transferring such short acoustic pulses into a three-dimensional crystal without increasing its wavelength has been a great challenge \cite{Temnov13}.
	
Optical generation of acoustic pulses in indirect band gap semiconductors GaP and Si without aid of metallic transducers has been studied recently in pump-probe reflectivity scheme \cite{Ishioka17}.  The generation of the acoustic pulses in these semiconductors originated from the short ($\sim$100 nm) absorption length of the pump light.  GaP can be grown with nearly perfect lattice match on exact Si(001) substrate, and the intermixing layer at the interface can be reduced to $\lesssim$7 atomic bilayers \cite{Beyer12, Beyer16}.  The combination of the two semiconductors therefore has the potential not only for Si based opto-electronic devices and high efficiency multi-junction solar cells but also for opto-acoustic transducers.   

In the present paper, we report on coherent acoustic phonons generated by excitation of the GaP/Si interface with a femtosecond laser pulse. Transient reflectivity responses exhibit a sharp spike at a time delay corresponding to the travel of the normal strain across the GaP layer, confirming its origin as an acoustic pulse generated at the GaP/Si interface and detected at the GaP surface and vice versa.  The spatial extent of the acoustic pulse can be directly estimated from the temporal width of the reflectivity spike.  The transient reflectivity responses also exhibit quasi-periodic Brillouin oscillations arising from the acoustic pulses during their propagation in the three-dimensional crystals.  Our theoretical modeling supports our interpretation that photoexcited carriers accumulated in the vicinity of the interface and surface give rise to the observed reflectivity changes.


The samples studied are nominally undoped GaP films grown by metal organic vapor phase epitaxy on $n$-type Si(001) substrates with different thicknesses from $d=$16 to 56 nm.   Details of the sample preparation and the evaluation of the atomic-scale structure are described elsewhere \cite{Beyer12}.  Transmission electron microscopy (TEM) images confirm that all the GaP layers have smooth surfaces and abrupt GaP/Si interfaces with intermixing of $\lesssim$7 atomic bilayers \cite{Beyer16}, while the density of planar defects is minimized.  The thicknesses of the GaP layers are measured by x-ray diffraction.  

Pump-probe reflectivity measurements are performed in a near back-reflection configuration using laser pulses with 400-nm wavelength (3.1-eV photon energy) and 10-fs duration \cite{Ishioka15}.  Pump-induced change in the reflectivity $\Delta R$ is measured as a function of time delay between pump and probe pulses using a fast scan technique.  The optical absorption depth, $\alpha_\textrm{GaP}^{-1}=$116 nm \cite{Aspnes83}, exceeds the GaP film thicknesses ($d$=56 nm at maximum).  The 3.1-eV photons are also absorbed strongly by the Si substrate ($\alpha_\textrm{Si}^{-1}=$82 nm).  


\begin{figure}
\includegraphics[width=0.475\textwidth]{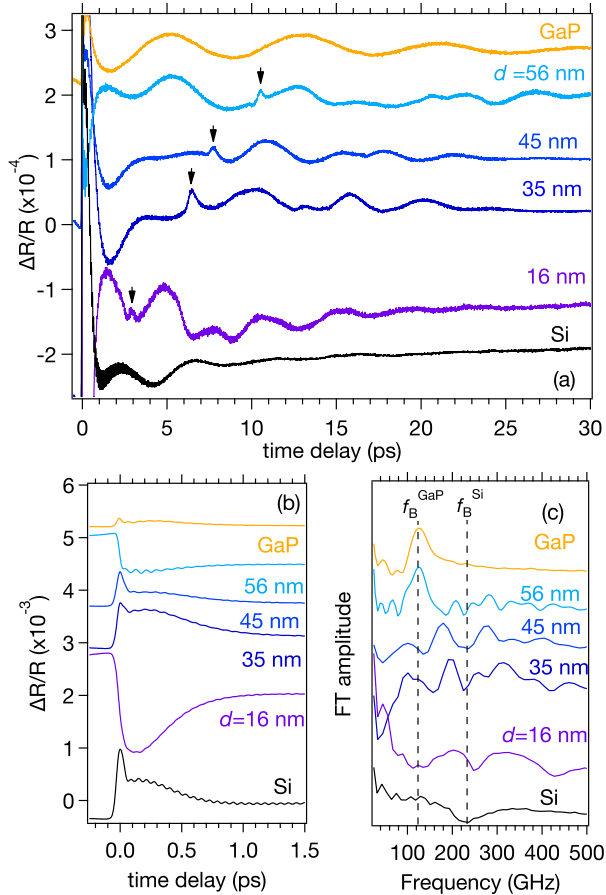}
\caption{\label{TD400} (Color Online.) (a,b) Reflectivity changes of the GaP films of different thicknesses $d$ on Si(001) substrates, together with those of bulk Si and GaP. Pump densities are 90$\mu$J/cm$^2$ for bulk Si and 30$\mu$J/cm$^2$ for other samples.  (a) and (b) show identical traces with different vertical and horizontal scales.  Arrows in (a) indicate the reflectivity spikes.  (c) Fourier-transformed spectra of the reflectivity changes for time delay $t>$0.1 ps.  }
\end{figure}

Figure~\ref{TD400}a compares the reflectivity responses from the GaP films of different thicknesses $d$ on Si, together with those of the bulk GaP and Si already reported in Ref. \cite{Ishioka17}.  All the traces show steep rise or drop at $t=$0, followed by recovery on a time scale of 1 ps (Fig.~\ref{TD400}b), which arises from the photoexcited carrier dynamics.  The reflectivity traces are also modulated by oscillations with sub-100 fs periods and several ps dephasing times due to the generation of coherent optical phonons of GaP and Si, whose details have been reported elsewhere \cite{Ishioka16}.  

On the longer time scale, the reflectivity traces of the GaP/Si samples show quasi-periodic oscillations on tens of picoseconds time scale, as shown in Fig.~\ref{TD400}a.  Similar but more regular oscillations are observed for bulk GaP and Si, and attributed to the interference between the probe beam reflected from the surface and also from the propagating strain pulse \cite{Ishioka17}.  The frequency $f_B$ of such an oscillation, sometimes referred to as Brillouin oscillation, is given by $f_B =2nv/\lambda$=123 and 235 GHz for GaP and Si for normal incidence, with $n$ being the refractive index, $v$, the longitudinal acoustic (LA) phonon velocity, and $\lambda$, the probe wavelength \cite{Thomsen86, Ishioka17}.  Fourier-transformed (FT) spectra for the thickest ($d=$56 nm) and thinnest ($d=$16 nm) GaP films (Fig.~\ref{TD400}c) are dominated by a peak at $f_B^\textrm{GaP}$ and a dip at $f_B^\textrm{Si}$, confirming an acoustic pulse propagating in the GaP film and in the Si substrate as the main origin of the reflectivity modulation, respectively.  We note that the $f_B^\textrm{Si}$ component appears as a dip because of the interference with the large non-oscillatory electronic response in the reflectivity, which we do not subtract before performing the Fourier transform.  The FT spectra for the intermediate thicknesses ($d=$35 and 45 nm), by contrast, have both $f_B^\textrm{GaP}$ and $f_B^\textrm{Si}$ components, indicating that both the acoustic pulses moving in GaP and in Si are contributing.  

\begin{figure}
\includegraphics[width=0.475\textwidth]{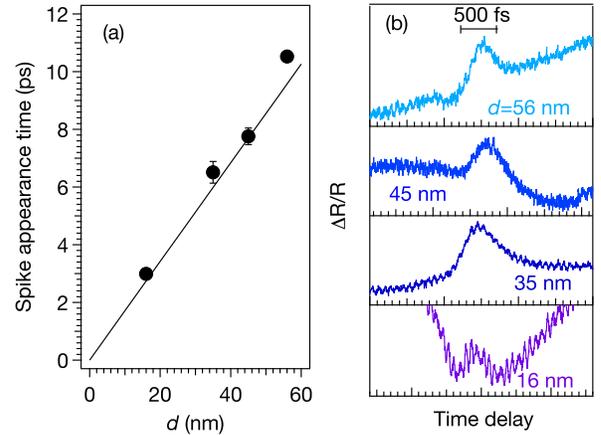}
\caption{\label{Blip} (Color Online.)  (a) Appearance time of the reflectivity spike as a function of the GaP film thickness $d$.  Solid line represents $t=\tau_0\equiv d/v^\textrm{GaP}$. (b) Expanded reflectivity traces showing the reflectivity spikes. }
\end{figure}

In addition to the quasi-periodic modulations, the reflectivity traces from the GaP/Si samples exhibit extra features seen as sharp spikes, as indicated by arrows in Fig.~\ref{TD400}a.  These spikes are absent for bulk GaP and Si and therefore characteristic of the interface.  The most intriguing feature of the spike is its extremely short ($\sim$0.5 ps) temporal width, as shown in Fig.~\ref{Blip}b.  The temporal width translates to a spatial extent of $\sim$3 nm or $\sim$10 atomic Ga-P bilayers, which is comparable with the thickness of the GaP-Si intermixing region at the interface, 7 atomic bilayers \cite{Beyer16}.  The spikes appear at delay times $t\simeq\tau_0\equiv d/v^\textrm{GaP}$, as shown in Fig.~\ref{Blip}a, with $v^\textrm{GaP}$ =5.847 nm/ps being the group velocity of the LA phonon of GaP in the [001] direction \cite{Weil68}.  The appearance times imply that the reflectivity spikes are induced by the LA phonon wave packet after traveling in the GaP film \emph{one way}, i.e., by acoustic pulses generated at the GaP/Si interface and detected at the GaP/air surface, and/or vice versa.    

\begin{figure}
\includegraphics[width=0.47\textwidth]{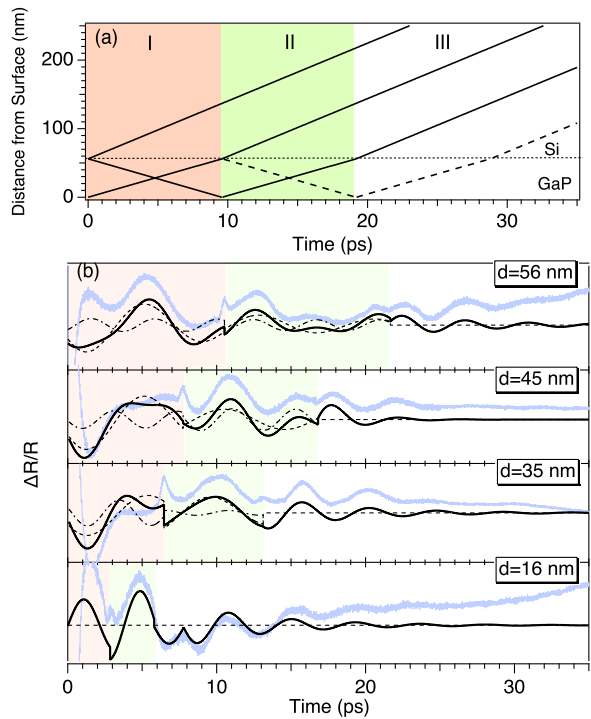}
\caption{\label{WA} (Color Online.) (a) Schematic diagram of the acoustic pulse positions as a function of time in the case of $d=$ 56 nm.  Dashed line represents the acoustic pulse position reflected at the interface.  (b) Fits of the experimental Brillouin oscillations to a double damped harmonic function.  Dashed, chained and solid curves represent the oscillation components at frequencies $f_B^\textrm{GaP}$ and $f_B^\textrm{Si}$ and their sum.  Thin blue curves represent the experimental traces.
}
\end{figure}

Figure~\ref{WA}a schematically shows the possible time evolution of the acoustic pulses. For $0<t\leq\tau_0$ (temporal regime I) the pulse generated at the GaP/Si interface propagates into both the GaP and Si layers, whereas that generated at the GaP surface moves into the GaP layer.  For $\tau_0<t\leq2\tau_0$ (regime II) the pulse from the surface mostly (76\% in amplitude) enters into the Si but a small portion (10\%) is reflected at the interface; meanwhile, the pulse from the interface is reflected at the GaP surface and propagates back to the interface.   For $t>2\tau_0$ (regime III) all the pulses propagate in the Si layer if we neglect the small reflection at the interface (dashed line in Fig.~\ref{WA}a).  
Based on these expectations we fit the experiments to the sum of two harmonic oscillations in regimes I and II with fixed frequencies at $f_B^\textrm{GaP}$ and $f_B^\textrm{Si}$ (neglecting the damping), and to a single damped harmonic oscillation at $f_B^\textrm{Si}$ in regime III.  The fits, shown in Fig.~\ref{WA}b, reproduce the experimental Brillouin oscillations well for $d\geq$35 nm.  We see that the oscillation component at $f_B^\textrm{Si}$ (chained curves in Fig.~\ref{WA}b) has distinct discontinuities at both $t = \tau_0$ and $2\tau_0$, confirming that acoustic pulses are indeed generated at both the surface and the interface.  For $d=$16 nm, however, we cannot determine the oscillation component at $f_B^\textrm{GaP}$ in regimes I and II, mainly because the fitting time window width $\tau_0$ is too small compared with $1/f_B^\textrm{GaP}$.  Moreover, we need to introduce an additional discontinuity at $t\sim3\tau_0$ to fit the experimental Brillouin oscillation, suggesting that contribution from the acoustic pulse reflected at the GaP/Si interface (dashed line in Fig.~\ref{WA}a) is not negligible for this sample.  
 
\begin{figure}
\centering
\includegraphics[width=0.475\textwidth]{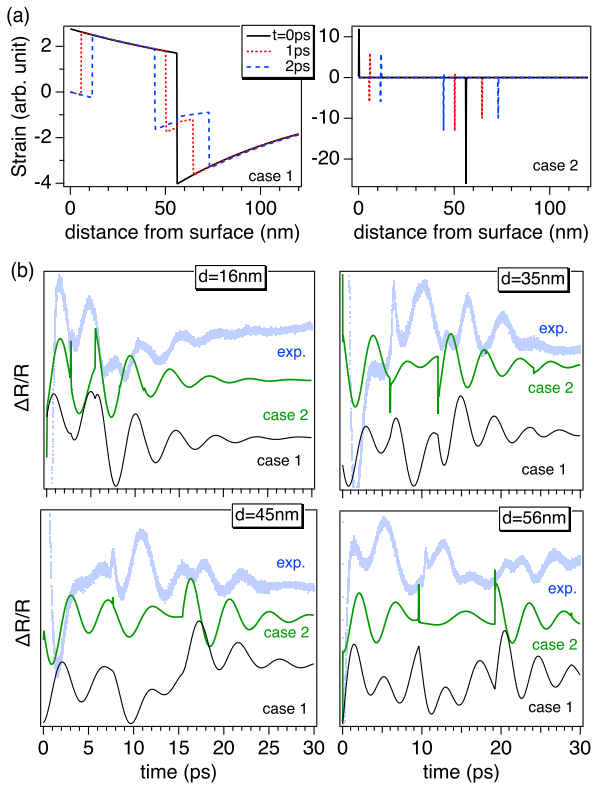}
\caption{\label{Ref_Th}  (Color Online.)  (a) 
(a) Time-dependent component of the carrier-induced strain at different times after photoexcitation for $d$=56 nm.   (b) Reflectivity changes in comparison with experiments for different GaP thicknesses $d$ on Si.  The calculations assume as-excited carrier distribution (case 1) and carriers localized at the interface and the surface (case 2).  The relative deformation potential coupling constants $a_{cv}^\textrm{GaP}$=2.25 eV and $a_{cv}^\textrm{Si}$=-4.52 eV are used for the both cases.   
Traces are offset for clarity.  }
\end{figure}

We have also theoretically modeled the optical generation and detection of an acoustic pulse at the GaP/Si heterointerface in the similar manner as those for bulk crystals \cite{Ishioka17}.  The details of the modeling are described in the Supplementary Material. 
We assume that a normal strain $\eta$ is generated predominantly via the deformation potential interaction with photoexcited carriers, and that the strain induces a change in the dependence of the dielectric constant $\epsilon$ on the probe photon energy $E$.  In this framework, both the generation and detection processes depend on the relative deformation potential coupling constant, $a_{cv}$, defined by the difference between the coupling constants of the conduction and valence bands.  
We take into account the contributions of the surface and interface displacements to $\Delta R/R$.
We first consider the \emph{as-excited} carrier density distribution $N(z)$ at distance from the surface $z$ that is proportional to the product of the absorption coefficient $\alpha(z)$ and the pump light intensity $I(z)$. In this case, $N$, and thereby the strain $\eta$, decay exponentially with $z$ and have a distinct kink at $z=d$ (case 1 in Fig.~\ref{Ref_Th}a and Fig. S1ab in Supplementary Material).  
This type of strain can reproduce the experimental Brillouin oscillations only qualitatively, and induce no sharp spike at $t=\tau_0$ in the reflectivity, regardless of the choice of $a_{cv}$ for GaP and Si, as demonstrated in as case 1 in Fig.~\ref{Ref_Th}b and Fig. S3 in Supplementary Material.

The failure of the \emph{as-excited} carrier distribution suggests that we need to take into account the redistribution of the photoexcited carriers occurring on shorter time scale than the acoustic pulse generation.  Indeed, the efficient generation of the coherent \emph{optical} phonons of GaP, which are observed simultaneously in the transient reflectivity changes (Fig.~\ref{TD400}b), indicates ultrafast separation of the photoexcited electrons and holes within $\sim$100 fs in the presence of space charge field (Fig. S1c in Supplementary Material) \cite{Ishioka16}.  Such ultrafast drift of photoexcited carriers, and their consequent accumulation at a heterointerface, can generate intense coherent LA phonons, as recently reported for the GaAs/transition metal oxide interface \cite{Pollock17}.  We model the carrier distribution $N$ after the charge separation with Gaussian functions centered near $z=0$ and $z=d$ (case 2 in Fig. S1a in Supplementary Material) and calculate the induced strain $\eta$ and the reflectivity change $\Delta R/R$ induced by the strain (case 2 in Fig.~\ref{Ref_Th} and in Figs. S1, S4 in Supplementary Material).  The calculations reproduce the experimental Brillouin oscillations only qualitatively also in this case.  However, they feature a distinct spike at $t=\tau_0$, though it appears either as a negative dip or a positive peak depending on $d$, in contrast with the always positive peak in the experiment.  The calculations support the sharp spike in the reflectivity arises from ultrashort strain pulses at the GaP/Si interface and the GaP surface, rather than the kinked exponential strain.

The most intuitive explanation for the reflectivity spikes would be the contribution from the surface and interface displacements when the acoustic pulses hit there.  Comparison between calculated $\Delta R/R$ with and without the surface and interface motions shows only minor difference  [Fig. S5 in Supplementary Material], however.  We therefore conclude that the spikes appearing in the calculated $\Delta R/R$ arise mainly from the discontinuity in the derivative of the signal at the boundaries and from the different deformation potentials in GaP and Si.  
Given that our calculations do not always reproduce the polarity of the experimental spikes, however, the latter may have a different origin.  For example, 
it is possible that  the photoexcited  carriers (electrons and holes) are carried by the acoustic pulse (acousto-electric effect) and modify the dielectric constant through the Drude contribution to the dielectric function when they reach the surface.
The discrepancy between the calculated and experimental Brillouin oscillations also indicates the limit of the present modeling, in which we use the same time-independent values of $a_{cv}^\textrm{GaP}$ and $a_{cv}^\textrm{Si}$ in describing both the generation and the detection.  Since both GaP and Si are indirect-gap semiconductors, the photoexcited carriers initially excited in the $\Gamma$ valley and along the $\Gamma-L$ valleys in respective semiconductor are scattered to the lower-lying $X$ valley on the time scale comparable with the acoustic phonon generation and detection, and the deformation potential coupling constants are expected to vary with time accordingly.  Whereas this can lead to a delayed build-up of the strain for the bulk crystals \cite{Ishioka17}, for the heterointerfaces the reflectivity response is far more complicated because the acoustic pulse can be generated in one material and detected in another.  
To include such dynamic effects in theoretical modeling, however, is beyond the scope of the present study.  


In conclusion, we have demonstrated that excitation of GaP/Si(001) heterointerfaces with femtosecond laser pulses induces ultrashort acoustic pulses, which are directly observed as sharp spikes in the transient reflectivity changes.  The sub-picosecond durations of the acoustic pulses are comparable with intermixing layer thickness at the GaP/Si interfaces, and considerably shorter than those reported in the previous studies.  
We thereby demonstrate that a simple optical pump-probe scheme enables us to nondestructively evaluate of the structural quality of the buried semiconductor heterointerface on nanometer scale through the direct observation of the acoustic pulse shape in the transient reflectivity.

\section*{Supplementary material}
See supplementary material for the details of theoretical simulations and their complete results.

\begin{acknowledgments}
This work is partly supported by the Deutsche Forschungsgemeinschaft through SFB 1083 and HO2295/8, as well as by NSF grant DMR-1311845 (Petek) and DMR-1311849 (Stanton).  
\end{acknowledgments}

\bibliographystyle{apsrev4-1}
\bibliography{AcousticInterface}

\end{document}